 \documentclass[12pt,preprint]{aastex}

\usepackage{graphics,color}
\definecolor{dark}{gray}{0.5}
\definecolor{red}{rgb}{1,0,0}
\definecolor{green}{rgb}{0,1,0}
\definecolor{blue}{rgb}{0,0,1}

\slugcomment{}

\shorttitle{Cluster in Dispersing Cloud}
\shortauthors{Chen \& Ko}

\begin{document}

\title{Indicators for cluster survivability in a dispersing cloud}

\author{Hui-Chen Chen\altaffilmark{1} and Chung-Ming Ko\altaffilmark{1,2}}
\affil{Institute of Astronomy, National Central University,
    Jhongli 320, Taiwan }

\altaffiltext{1}{Institute of Astronomy, National Central University,
    Jhongli 320, Taiwan}
\altaffiltext{2}{Department of Physics and Center for Complex Systems,
National Central University, Jhongli 320, Taiwan}

\begin{abstract}
We use N-body simulations to survey the response of embedded star clusters to the
dispersal of their parent molecular cloud.
The final stages of the clusters can be divided into three classes: the cluster
(i) is destroyed, (ii) has a loose structure, and (iii) has a compact core.
We are interested in three of the governing parameters of the parent cloud:
(i) the mass, (ii) the size, and (iii) the dispersing rate.
It is known that the final stage of the cluster is well correlated with the star formation
efficiency (SFE) for systems with the same cluster and cloud profile.
We deem that the SFE alone is not enough to address systems with clouds of
different sizes.
Our result shows that the initial cluster-cloud mass ratio at a certain Lagrangian radius,
and the initial kinetic energy are better indicators for the
survivability of embedded clusters.

\end{abstract}


\section{Introduction}

Stars are the fundamental units to build up a galaxy and are mostly not born alone but in groups.
Stellar groups, clusters, or associations are formed inside molecular clouds.
They are embedded and not optically seen until they get rid of the residual material after
star formation.
An embedded cluster is not always understood clearly because of observational constraints.
The near-infrared all sky survey, Two Micron All Sky Survey, has given astronomers a chance to look into the clouds.
\citet{lad03} claimed that the number of embedded clusters declines rapidly with age and
the infant mortality of clusters (i.e., disruption not long after birth)
is more than 90\% in our Galaxy.

Galactic tidal forces, close encounters with giant molecular clouds
\citep{gie06}, shock heating, and
mass loss of the system by massive member stars,
such as UV radiation, stellar winds, and supernova explosions \citep{boi03a,boi03b},
are possible disruption mechanisms for stellar groups, whether they are embedded or not.
Nonetheless,most of these mechanisms (such as the first three aforementioned mechanisms) have a
destruction timescale longer than the upper limit of the lifetime of molecular clouds, which
is about a few to a few tens of Myr \citep{bli80,elm00,har01,bon06}.
CO observations in our Galaxy suggested that
the lifetime of molecular clouds is of the order of 10 Myr \citep{lei89}.
Timescale estimation of photoevaporation and statistics
on the expected numbers of stars per cloud showed that giant
molecular clouds of mass $10^6 M_{\odot}$ are expected to survive
for about 30 Myr \citep{wil97}.
The most promising disruption mechanism for embedded clusters
seems to be the dispersion of the parent cloud by UV radiation,
stellar winds, and supernova explosions in early cluster evolution
\citep[see, e.g.,][]{tut78,lad84,goo97,boi03b,bau07,bas06,goo06}.

In a large set of simulations, \cite{bau07} studied the dispersal of the residual
gas by decreasing the mass with different star formation efficiency (SFE),
and in different tidal fields.
They concluded that the clusters had to form with SFE $\geq$ 30\%
in order to survive gas expulsion, and the external tidal fields have significant
influences only if the ratio of half mass radius to tidal radius is larger than 0.05.
\cite{goo06} and \cite{bas06} addressed a similar problem and found that the embedded
clusters would be destroyed within a few tens of Myr if the ``effective star formation
efficiency'' (hereafter eSFE) $\leq$ 30\%.

\citet{che08} studied the behavior of embedded clusters when the parent cloud is dispersing
(i.e., the size increases but the total mass remains constant, thus the cloud becomes more and
more diluted).
From a large survey of simulations we found that the final structure can be classified,
according to the expansion ratio of $r_{45}$ (the 45\% Lagrangian radius of the cluster),
into three groups: (i) destroyed, (ii) loose structure, and (iii) compact core.
Empirically, the expansion ratio is related to the cluster-cloud mass ratio (which will be
defined later in this paper).

In this work, we address the problem of what determines the fate of the embedded cluster.
We consider that the cluster and the parent cloud could have different initial density profiles.
Hence we examine three parameters of the parent cloud independently: (i) mass, (ii) size, and
(iii) dispersing rate of the cloud.
Clusters with and without mass functions are considered.
The paper is organized as follows.
In Section~2, we describe the model and simulations.
In Secttion~3, we discuss the applicability of SFE as an indicator
for describing the behavior of the cluster, and then introduce other better indicators for
more general initial conditions.
A summary and some remarks are provided in Section~4.

\section{Model}\label{sec:model}
To study the survivability of embedded star clusters, we consider an idealized model.
We put a spherical star cluster at the center of an expanding spherical molecular cloud,
and then examine the subsequent behavior numerically.
In our simulations, the distribution of stars in the cluster is \citet{plu11}
with a length scale of 0.6 pc (note that the half mass radius is $\sim 0.8$ pc).
The cloud is represented by an external \citet{plu11} potential in which
the length scale increases with time to represent the dispersion of a cloud.
We adopt the $N$-body simulation code $NBODY2$ \citep{aar01}
to study the evolution of the clusters.

Similar systematic simulations have been performed by, e.g.,
\citet{gey01} and \citet{bau07}.
\citet{gey01} used the \citet{kin66} model to described both the cloud and gas, and
\citet{bau07} used the \citet{plu11} model.
Both papers considered equal-mass stars and represented the cloud dispersion as
a continuous reduction of cloud mass.
In contrast, we mimic cloud dispersion as the expansion of the cloud and the cloud
mass does not change.

\subsection{Initial condition}
We assume that each embedded cluster contains 2500 stars with a total mass of
2500 $M_{\odot}$.
The physical distribution is Plummer and the length scale, $a_c$, is about 0.6 pc.
The velocity distribution depends on the potential of the parent cloud.
The initial mass function (IMF) with \citet{sal55} slope from 0.3 to 30 $M_{\odot}$
is also considered, and there is no primordial mass segregation.
All clusters, with the parent clouds, are required to be in virial equilibrium
prior to cloud dispersion.

The SFE
(and more precisely, the total SFE)
is defined as
\begin{equation}\label{eta_definition}
\eta=\frac{M_{c}}{{M_{c}+M_{b}}}\,,
\end{equation}
where $M_c$ is the mass of the cluster and $M_b$ is the mass of the parent cloud.
In our runs $M_b$ ranges from 0.5 to 19 $M_c$, which gives $\eta$ from 67\% to 5\%.

\citet{goo08} also discussed cases of a cluster and cloud with different spatial
distributions but the cluster is in virial equilibrium initially.
This is very similar to how we set up our initial conditions.
Note that the eSFE of \citet{goo08} is related to the length scale of the cloud.
More discussion on eSFE will come later.

\subsection{Cloud dispersion}
The cloud is represented by an external Plummer potential,
\begin{equation}
\Phi_{P}=\frac{-GM_b}{\sqrt{r^2+a^2}}\,,
\end{equation}
where $G$ is the gravitational constant, $r$ is the distance from the cloud center
(which is also the star cluster center), and $a$ is the length scale of the potential.
To mimic expansion, we let $a$ increases with time,
\begin{equation}
a=a_0 {e}^{t/t_e}\,,
\end{equation}
where $a_0$ is the initial length scale and $t_e$ is the dispersing e-fold timescale.
A molecular cloud has a relatively short lifetime ranging from
a few to a few tens of Myr.
We try four different e-fold timescales: from 0.33 to 3.3 Myr.
The potential becomes ineffective in no more than two e-fold times
(largest dispersing timescale) to 10 e-fold times (smallest dispersing timescale),
which corresponds to 7-3 Myr.
As molecular clouds are not expected to last longer than 30 Myr, we stop our simulations
at 30 Myr after the cloud starts dispersing.
In any case, the influence of the gas removal is long gone well before the end of the
simulations.

\section{Results and discussion}
We assume that the embedded clusters are bound and in virial equilibrium initially.
When the cloud disperses, the cluster starts to expand.
It may be destroyed or may shrink back later by self gravity.

\subsection{Cluster and cloud have the same initial density profile}
\begin{figure}[htbp]
\includegraphics[width=0.45\textwidth, angle=0]{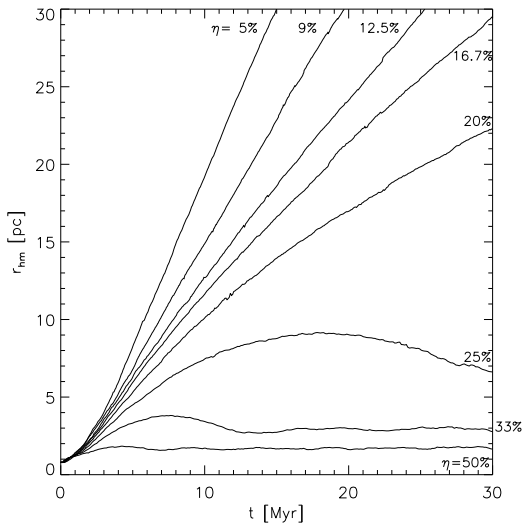}
\caption{
        Evolution of half-mass radius as a function of time. Here, the dispersing
        timescale $t_e$ is 1.1 Myr, SFE is from 5\% to 50\%.
         }
\label{lagr_eta_0.3}
\end{figure}

In this subsection, we present some cases for clouds having the same Plummer length scale
as the clusters do, 0.6 pc.
In these cases, SFE is a good indicator for the subsequent behavior of the clusters.
Figure~\ref{lagr_eta_0.3} shows the half mass radius ($r_{hm}$) evolution in different
SFEs with a cloud dispersing (or gas removal)
timescale of 1.1 Myr. $r_{hm}$ increases
almost linearly when SFE is less than 20\%. Clusters are destroyed in these cases.
For cases with SFE from 25\% to 50\%, $r_{hm}$ expands slower and shrinks back
(even becomes stable) later. In these runs, clusters remain intact but
with different concentrations.
The expansion of a cluster by gas removal and then shrinking (collapse) to equilibrium later was
also reported in the simulations by \citet{goo06} and \citet{bau07},
and has also been compared with observations by \citet{bas08}.

\subsubsection{Different dispersing timescales}
\begin{figure}[htbp]
\includegraphics[width=0.45\textwidth, angle=0]{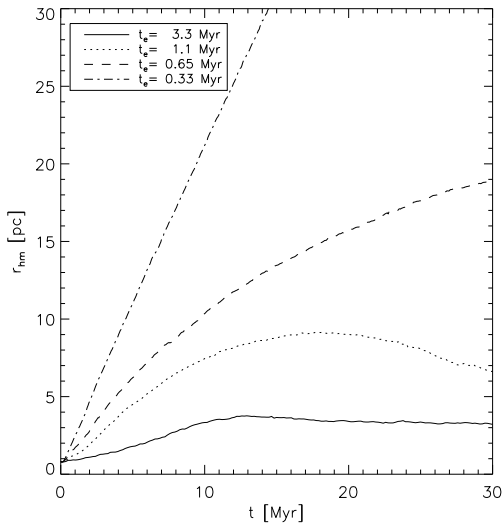}
\caption{
        Evolution of half-mass radius as a function of time, in terms of Myr.
        The dispersing timescale here is from 0.33 to 3.3 Myr. The SFE is 25\%.
         }
\label{lagr_dispersing}
\end{figure}

The behavior is expected to be different if we change the cloud dispersing timescale
(or gas removal timescale).
Figure~\ref{lagr_dispersing} shows
the $r_{hm}$ evolution of SFE $=$ 25\% with four dispersing timescales.
$r_{hm}$ increases almost linearly when $t_e$ is 0.33 Myr, and increases slower
when $t_e$ is 0.65 Myr.
For $t_e$ equal to 1.1 and 3.3 Myr, $r_{hm}$ increases much more slowly and changes little
after some time.

\subsubsection{Expansion ratio of $r_{hm}$ and bound mass fraction}
After dispersing, all clusters expand.
To compare the structure before and after, we define the expansion ratio
of the half-mass radius $r_{hm}$ as
\begin{equation}\label{eps_definition}
\epsilon={r_{hm}({\rm final})\over r_{hm}({\rm initial)}}\,.
\end{equation}
Here final means 30 Myr.

\begin{figure}[htbp]
\includegraphics[width=0.45\textwidth, angle=0]{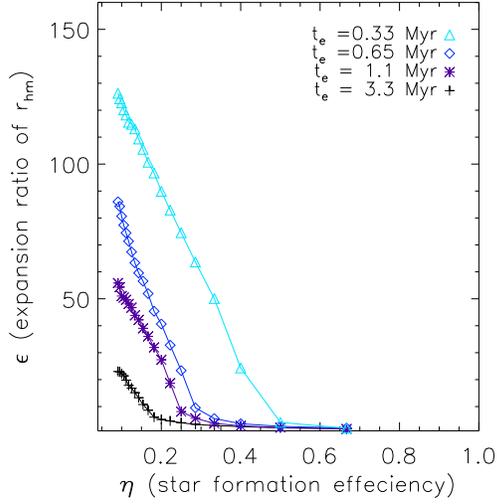}
\caption{
        Expansion ratio of half-mass radius as a function of SFE.
        (A color version of this figure is available in the online journal.)
         }
\label{eps_rhm}
\end{figure}

Figure~\ref{eps_rhm} represents the results for the expansion ratio of $r_{hm}$
at the end of the simulations, 30 Myr, as a function of SFE in different dispersing timescales.
Different dispersing timescales produce different expansion ratios for the same SFE.
For each dispersing timescale, there is a simple relation between $\eta$ and $\epsilon$.
Each relation has two branches jointed by a turnover point.
Specifically, the turnover points for $t_e= 3.3, 1.1, 0.65, 0.33$ Myr are at
$\eta\approx 0.18, 0.25, 0.33, 0.50$, respectively.
The steeper branch corresponds to destroyed clusters and the flatter branch to survived clusters.
Hence the SFE, $\eta$, is a good indicator for the survivability of embedded clusters.

\begin{figure}[htbp]
\includegraphics[width=0.45\textwidth, angle=0]{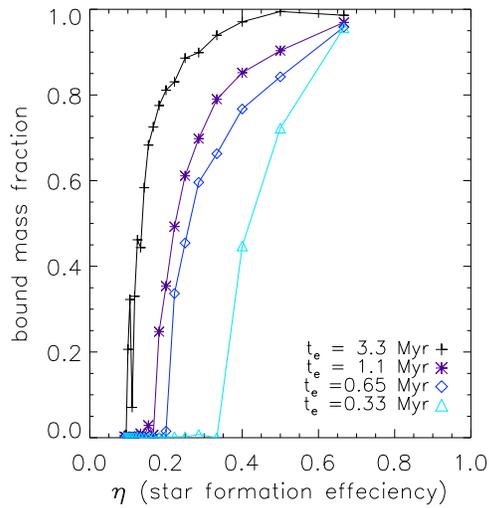}
\caption{
        Bound mass fraction as a function of SFE.
        (A color version of this figure is available in the online journal.)
         }
\label{boundmass}
\end{figure}

Figure~\ref{boundmass} depicts the bound mass fraction as a function of SFE in
different dispersing timescales (or gas removal timescale).
Both Figures~\ref{eps_rhm} and \ref{boundmass} nicely demonstrate the expected result:
a cluster might survive in a slowly dispersing cloud but might be destroyed in
a fast dispersing one.
The dependence of the bound mass fraction on SFE is also reported by \citet{bau07}.
Based on this relation, \citet{par08} discussed the shape of the initial cluster mass function.

\subsection{Cluster and cloud have different initial density profiles}
\begin{figure}[htbp]
\includegraphics[width=0.45\textwidth, angle=0]{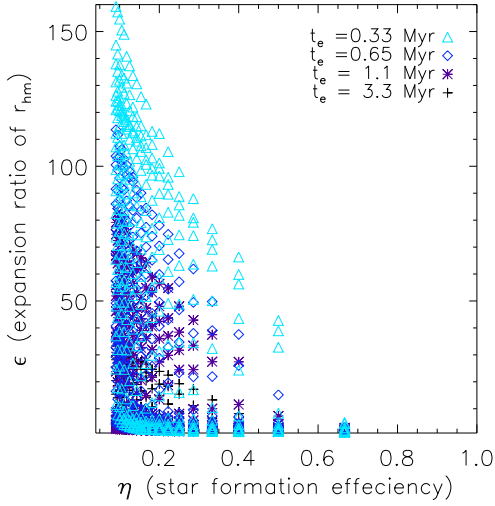}
\caption{
        A plot of the expansion ratio of half-mass radius against SFE for 20 cases of $a_0$. 
        (A color version of this figure is available in the online journal.)
         }
\label{eps_rhm_eta}
\end{figure}

We suppose that it is too idealized to consider both cluster and cloud have the same initial
density profile (as in the previous section), and it is conceivable that they are more likely to
have different initial profiles.
To study the cases with different initial profiles, we fixed the Plummer length scale of
the cluster at 0.6 pc, but varied the Plummer length scale of the cloud from four times smaller to
four times larger than that of the cluster.
Thus, there are two parameters to describe the initial profile of the cloud: $\{M_b,a_0\}$,
its mass and the initial Plummer length scale. (One can use the SFE $\eta$ and $a_0$ instead.)
It is obvious that $\eta$ alone is not sufficient to describe the results with different
$a_0$ and $M_b$ (or equivalently $\eta$, see Equation~(\ref{eta_definition})).
When we plot the expansion ratio $\epsilon$ of 20 $a_0$ cases against $\eta$
(see Figure~\ref{eps_rhm_eta}), no discernible relation(s) can be found.
(We should point out that if we plot the result of one $a_0$, the graph will be similar to
Figure~\ref{eps_rhm}.)

\begin{figure}[htbp]
\includegraphics[width=0.45\textwidth, angle=0]{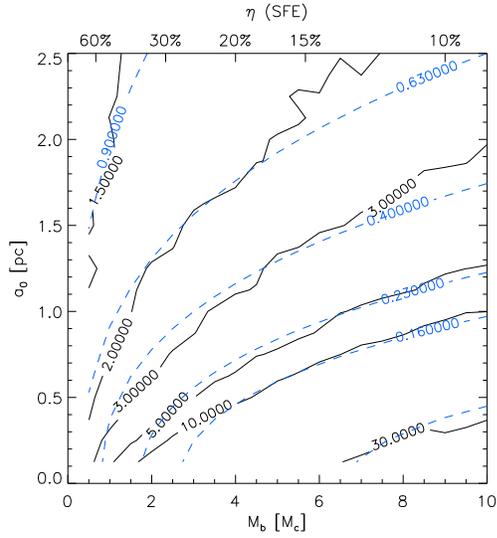}
\caption{
        Contours of the expansion ratio of half-mass radius
        ($\epsilon_{r_{hm}}=\epsilon_{r_{50}}$, solid line) in the parameter space
        $\{M_b,a_0\}$. The dashed lines are contours of the cluster-cloud mass ratio
        at half-mass radius, $\beta_{r_{hm}}=\beta_{r_{50}}$.
        In this example, the cloud dispersing timescale is $t_e=3.3$ Myr.
        The corresponding SFE is labeled at the top.
        (A color version of this figure is available in the online journal.)
        }
\label{eps_mb_a0}
\end{figure}

However, $\epsilon$ does show a ``simple relation'' with $\{M_b,a_0\}$.
Figure~\ref{eps_mb_a0} shows the contours of $\epsilon$ in the parameter space $\{M_b,a_0\}$
for the dispersing timescale $t_e=3.3$ Myr.
Understandably, $\epsilon$ is large/small (cluster is destroyed/remains intact)
when $a_0$ is small/large and $M_b$ is large/small (or $\eta$ is small/large).
In fact, there are three final configurations for the cluster:
(i) destroyed, (ii) loose structure, and (iii) compact core.
The boundaries between these configurations closely match the contours $\epsilon=10$ and $2$
for the cases with different dispersing rates \citep[for details, see][]{che08}.
In other words, on the one hand for a fixed effective size of the natal cloud,
it is clear that the survivability of the cluster is higher when the SFE is larger.
On the other hand, for a fixed SFE, the survivability increases when the effective
size of the natal cloud increases.
For example, when the cloud contains 90\% of the mass of the system (i.e., total SFE 10\%),
the half-mass radius of the cluster expands at least twice in 30 Myr.
The smaller the size of natal cloud, the more the star cluster will expand.
Apparently, the reason is the dynamics of the cluster is dominated more by the cloud
when the mass of cloud initially enclosed within the cluster is more,
such as the cases of small SFE and small effective natal cloud size.
And in contrast for the cases of large SFE and large effective natal cloud size,
the dynamics is dominated by the cluster.

Although SFE $\eta$ fails to be a good survivability indicator of the two-parameter system,
we are able to identify other survivability indicators: the initial cluster-cloud mass ratio
and the initial kinetic energy. We describe these two indicators in the following separately.

\subsubsubsection{Cluster-cloud mass ratio}

First denote the mass fraction $f$ Lagrangian radius of the cluster as $r_f$,
i.e., within $r_f$ the mass of the cluster is $fM_c$, and denote this mass by $M_{c,r_f}=fM_c$.
Also denote $M_{b,r_f}$ as the mass of the cloud within $r_f$ initially.
Define the cluster-cloud mass ratio as \citep[see][]{che08}
\begin{equation}\label{beta}
\beta_{r_f}=\frac{M_{c,r_f}}{M_{c,r_f}+M_{b,r_f}}\,.
\end{equation}
This closely resembles the definition of SFE $\eta$ (see Equation~(\ref{eta_definition})),
but $\beta_{r_f}$ contains information about the length scale of the cloud $a_0$.
Note that $\beta_\infty=\eta$ (when $f=1$, $r_f\rightarrow\infty$).
Since we are using the Plummer model for both cluster and cloud, there is a simple relation
between $\beta_{r_f}$ and $\eta$
\begin{equation}
\left({1\over\eta}-1\right)={f\over F}\left({1\over\beta_{r_f}}-1\right)\,,
\end{equation}
where $f=M_{c,r_f}/M_c$ and $F=M_{b,r_f}/M_b=r_f^3/(r_f^2+a_0^2)^{3/2}$.
We also extend the half-mass radius expansion ratio of Equation~(\ref{eps_definition}) to
\begin{equation}\label{epsf_definition}
\epsilon_{r_f}={r_f({\rm final})\over r_f({\rm initial)}}\,.
\end{equation}

\begin{figure}[htbp]
\includegraphics[width=0.45\textwidth, angle=0]{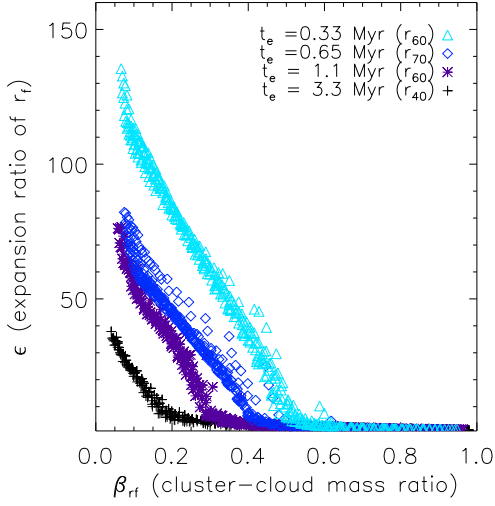}
\caption{
     Expansion ratio of the chosen Lagrangian radii as a function of the cluster-cloud mass,
     $\beta_{r_f}$.
     (A color version of this figure is available in the online journal.)
         }
\label{eps_r_beta}
\end{figure}

We find that for each dispersing timescale $t_e$, $f$ can be tuned such that data from the
two-dimentional parameter space $\{M_b,a_0\}$ collapses to a relation in the $\beta_{r_f}-\epsilon_{r_f}$
plane, see Figure~\ref{eps_r_beta}.
The relation closely resembles the relation $\eta-\epsilon$ for cases of fixed $a_0$
(compared to Figures~\ref{eps_r_beta} and \ref{eps_rhm_eta})  and the turnover points
for $t_e= 3.3, 1.1, 0.65, 0.33$ Myr are at $\beta_{r_f}\approx 0.18, 0.28, 0.40, 0.53$,
respectively.
We note that $r_f$ does not increase monotonically with increasing dispersion timescale.

\begin{figure}[htbp]
\includegraphics[width=0.45\textwidth, angle=0]{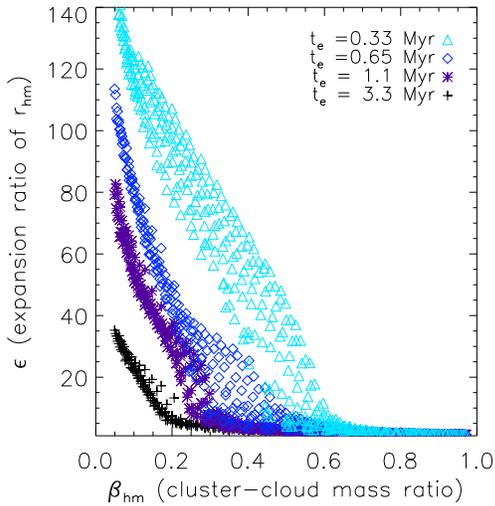}
\caption{
     Expansion ratio of the half-mass radius as a function of the cluster-cloud mass,
     $\beta_{r_f}$. Here $f$ is $0.5$.
     (A color version of this figure is available in the online journal.)
         }
\label{eps_rhm_beta}
\end{figure}

We stress that although it may seem a bit contrived that we have to tune $r_f$ to obtain a tight
relation, the result is reasonably good even if we pick one particular $r_f$ for all timescales.
Figure~\ref{eps_rhm_beta} shows the relations at half-mass radius for all dispersing timescales
(i.e., $\epsilon_{r_{hm}}$ against $\beta_{r_{hm}}$).
In Figure~\ref{eps_mb_a0}, we see that the contours of $\epsilon_{r_{hm}}$  and $\beta_{r_{hm}}$
agree with each other in the parameter space $\{M_b,a_0\}$.

\subsubsubsection{Initial kinetic energy or virial energy}

\begin{figure}[htbp]
\includegraphics[width=0.45\textwidth, angle=0]{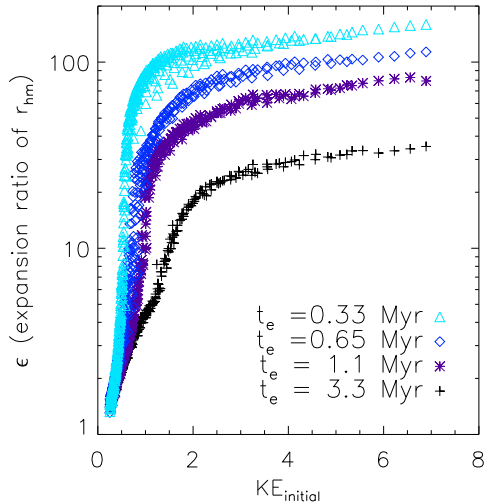}
\caption{
  Initial kinetic energy and the expansion ratio
  of half-mass radius. The units of the initial kinetic energy here are
  in $N-body$ units.
  (A color version of this figure is available in the online journal.)
         }
\label{ki_eps}
\end{figure}

To clarify the relation between the final state and the initial condition,
we also check the energies.
We find that the initial kinetic energy is nicely correlated with
the expansion ratio of half-mass radius $\epsilon$, see Figure~\ref{ki_eps}.
Since the initial system is in virial equilibrium,
the initial virial energy is just twice of the kinetic energy,
hence it should also be a good indicator.

The initial kinetic energy and the initial virial energy show
how the cloud mass affect stars inside the cluster.
$\beta_{r_f}$ is the cluster-cloud mass ratio within a certain region.
All of them show nice relations with the expansion ratio $\epsilon$ or $\epsilon_{r_f}$.
But if one would like to apply to observations, it might not be easy
to obtain the energy estimation for the whole cluster.
$\beta_{r_f}$ should be the more convenient choice.

\begin{figure}[htbp]
\includegraphics[width=0.85\textwidth, angle=0]{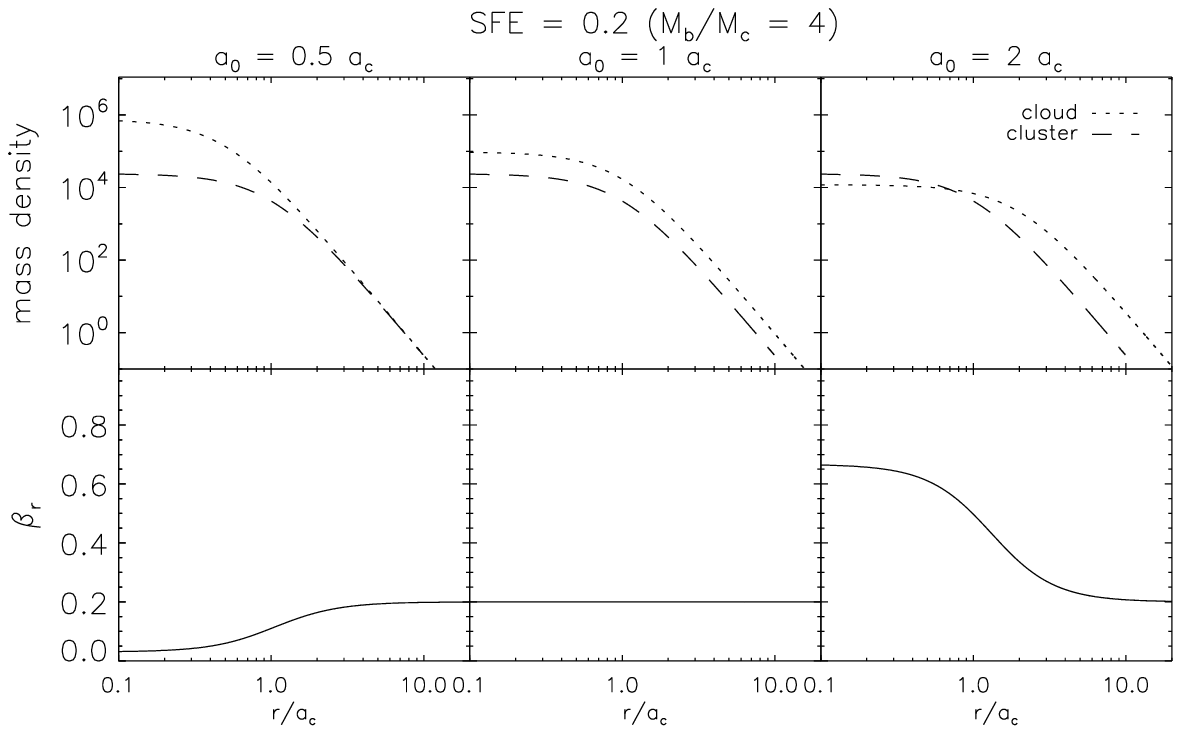}
\caption{
  Mass density profiles and the corresponding cluster-cloud mass ratio, $\beta_r$, for cases
  with different length scales of the cloud. The total SFE $\eta=0.2$ (i.e., $M_b/M_c=4$).
  From left to right, the ratios of length scale of the cloud to cluster are 0.5, 1, and 2,
  respectively. The mass of cloud is more concentrated (cloud-dominated) in the left column
  and more spread out (cluster-dominated) in the right column. The mass densities are in arbitrary units.
         }
\label{rho_beta}
\end{figure}

Figure~\ref{rho_beta} shows three initial mass density profiles of clusters and clouds at a
particular cloud mass $M_b/M_c$ or total SFE $\eta$. In the figure, $M_b/M_c=4$ and $\eta=0.2$.
The upper panels show the density profiles against radius and the lower panels show
the corresponding $\beta_r$ against radius.
Note that $\beta_r\rightarrow\eta$ as $r\rightarrow\infty$.
In the middle column, $a_0=a_c$, the two density distributions are the same, and
the corresponding $\beta_r$ is always 0.2 at any radius.
In the left column, $a_0<a_c$ (i.e., comparatively the cloud mass is distributed more
concentrated at the origin than the cluster),
$\beta_r$ increases monotonically with $r$ (from $\beta_r<\eta$ towards $\eta$ as
$r\rightarrow\infty$ and $\beta_r$ always smaller than $\eta$).
In the right column, $a_0>a_c$ (i.e., comparatively the cloud mass is distributed more
evenly than the cluster),
$\beta_r$ decreases monotonically with $r$ (from $\beta_r>\eta$ towards $\eta$ as
$r\rightarrow\infty$ and $\beta_r$ always larger than $\eta$).
Therefore, for every total SFE, we have three distinct types: $a_0$ $<$, $=$, $>$ $a_c$.
The cluster has a higher/lower chance to survive when the cloud size is larger/smaller
than the cluster size at a given total SFE.
This also suggested that at a low total SFE the cluster has a higher chance to survive if the
length scale of cloud is larger.

To this end, we would like to point out a similar idea introduced by \citet{goo08}.
The effective SFE, eSFE, is defined as the virial ratio of the stars after instantaneous
gas expulsion.
The total SFE equals the eSFE when the cluster and cloud share the same density profile.
When the cluster is born dynamically ``cold/hot'',
the velocity of a cluster is smaller/larger than what is required for virial equilibrium,
the eSFE is larger/smaller and the cluster has a better/worse chance to survive.
This is comparable to our results of clouds with larger/smaller length scales.

Moreover, eSFE is closely related to the initial kinetic energy of the cluster
that we mentioned earlier.
For a fixed total SFE, the initial kinetic energy is smaller if the
size of the natal cloud is larger.
In this case the cluster is `colder' and tends to survive after the cloud dispersed.

\subsection{Mass segregation}
\begin{figure}[htbp]
\includegraphics[width=0.45\textwidth, angle=0]{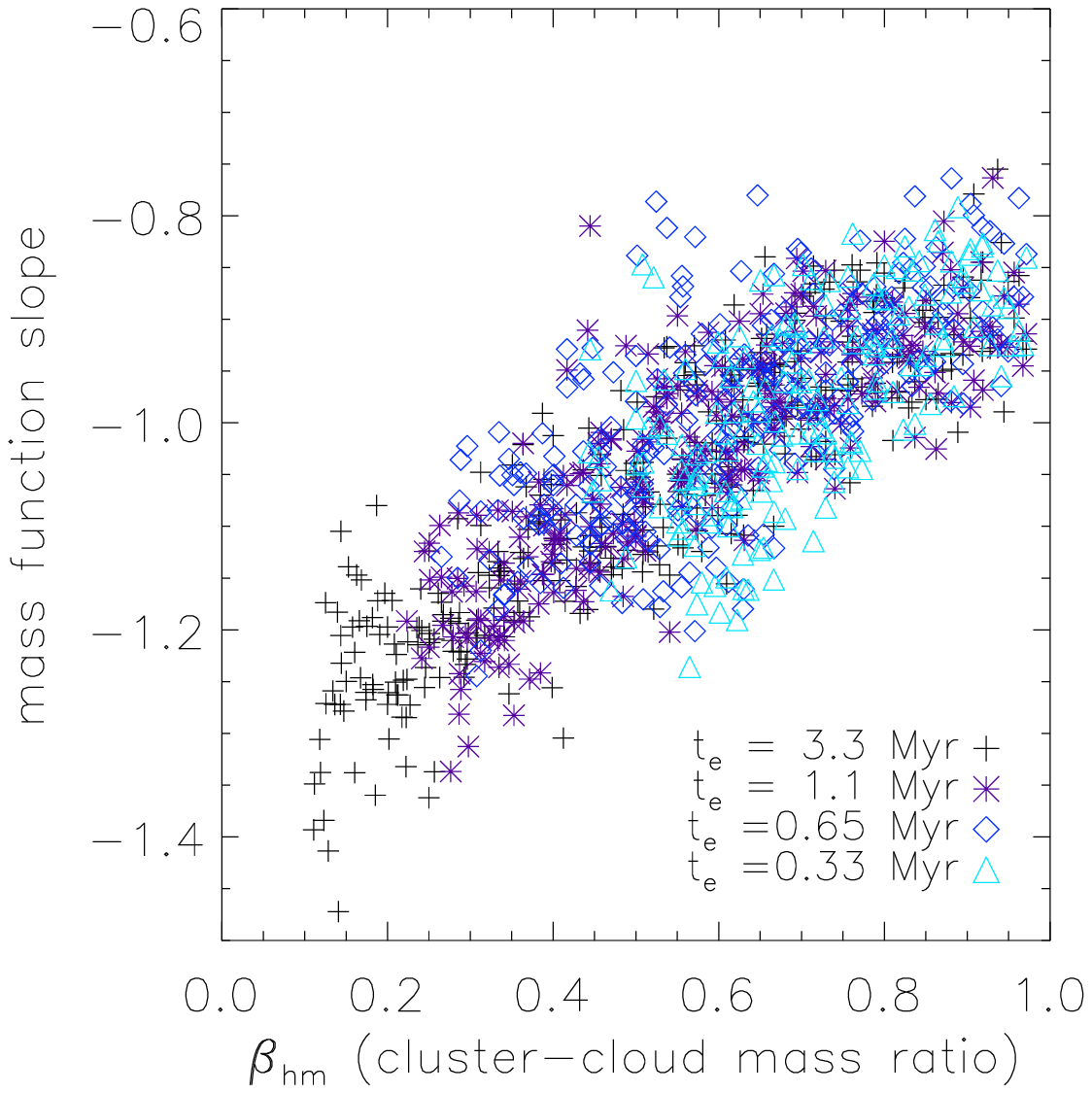}
\caption{
     Mass function slope as a function of half-mass $\beta$ value.
     The mass function slope is derived from stars within the final half-density (2D) radius.
     The IMF slope is $-1.35$ (Salpeter) for all cases.
     (A color version of this figure is available in the online journal.)
      }
\label{mfs_beta}
\end{figure}

Mass segregation is another important issue in stellar clusters.
Once again we use NBODY2 to study mass segregation.
All simulations start with the Salpeter mass function,
i.e., the mass function slope is $-1.35$.
We find that $\beta_{r_{hm}}$ is also a good indicator for the level of mass segregation.
Figure~\ref{mfs_beta} shows the mass function slope derived from within the
final half density radius as a function of $\beta_{r_{hm}}$ for cases
with $\epsilon_{r_{hm}} \le 20$.
When $\beta_{r_{hm}}$ is small (i.e., cloud mass is the dominant factor), the mass segregation
of the cluster is less severe.
We note that using different $\beta_{r_f}$ does not change the relation much.

The preferential loss of low-mass stars is due to two-body relaxation
\citep[see][]{spi58a,spi58b} and would flatten the mass function.
We should point out that we fit the mass function by a simple single power law,
which is not the best fit for all cases.
As a matter of fact, the broken power law observed in the Arches cluster \citep{kim06} is also
seen in our simulations.
We note that this could also be obtained by dynamical evolution without a natal cloud
\citep[see also][]{por07}.
However, we cannot find an obvious relation between the broken power law and SFE $\eta$ or the
cluster-cloud mass ratio $\beta$.
This may be due to poor statistics in our runs (2500 stars initially).
Moreover, the broad distribution of the final mass function slope in Figure~\ref{mfs_beta}
is perhaps a result of the simple single power law fitting.

\section{Summary and Remarks}
We studied the consequence for clusters after parent clouds disperse in
different dispersing timescale, cloud mass, initial cloud length scale, and IMF.
We mimic cloud dispersal by increasing cloud size or length scale, but keeping
the total cloud mass unchanged.

SFE, $\eta$, is a good indicator for the survivability of embedded cluster after
the dispersal of the parent cloud only if the cluster and cloud have the same density profile.
It fails badly if the cluster and cloud have different density profiles.
The latter cases can be described by two parameters $\{M_b,a_0\}$, the mass of the cloud
and the length scale of the Plummer model.
The expansion ratio of the cluster is obviously depending on the two parameters.
Although we cannot use $\eta$ as a survival indicator, we find two quantities that can serve
the same purpose.

The two quantities are (i) the initial cluster-cloud mass ratio (within a certain Lagrangian
radius of the cluster), $\beta_{r_f}$ (see Equation~(\ref{beta})), and 
(ii) the initial kinetic energy (or virial energy).
Each of them can organize the data of itself and the expansion ratio
(of a certain Lagrangian radius of the cluster) at a dispersing timescale
into a `one-dimensional' relation (effectively collapsing data from a surface into a line).
Moreover, the relation is very nice in the sense that it has two branches, one corresponds to
cluster destruction and the other to survival.
Hence the two quantities are good indicators for the survivability of embedded clusters.

We should point out that for different cloud dispersing timescales we need to choose
somewhat different Lagrangian radius to obtain the tightest correlation between
the cluster-cloud mass ratio and the expansion ratio.
Having said that, in fact, $\beta_{r_{hm}}$ fits the data reasonably well
for all different cloud dispersion rates .
The tight correlation strongly suggested that the evolution of the
embedded cluster depends on the total mass enclosed within the
corresponding Lagrangian radius of the cluster.
We thus propose that the relevant (initial) effective timescales should be defined in terms of the total mass
enclosed within the best-fit Lagrangian radius (or roughly the half-mass radius where $f=0.5$)
\citep[e.g.,][]{bin87},

\begin{equation}\label{timescales}
  \tau_{\rm relax}\approx {f N\over 8\beta_{r_f}^2\log(fN/\beta_{r_f})}\,\tau_{\rm cross}\,,
  \quad \tau_{\rm cross}\approx\sqrt{\beta_{r_f}r_f^3\over fGM_c}\,,
\end{equation}
where $N$ is the number of stars in the cluster, and
$\tau_{\rm relax}$ and $\tau_{\rm cross}$ are, respectively, the effective relaxation time and crossing time.

Furthermore, for clusters with Salpeter mass function, the slope of the final mass function
shows a roughly linear relation with the cluster-cloud mass ratio at half-mass radius,
$\beta_{r_{hm}}$, see Figure~\ref{mfs_beta}.
In addition, all the dispersing rates we considered share the same relation.

As mentioned in Section~\ref{sec:model}, similar works have been done with
an equal-mass model \citep[e.g.,][]{gey01,bau07}.
\citet{goo97} mentioned that the survivability of the clusters
is only slightly higher in the cases
of equal mass than cases with mass function.
For comparison with the result of mass function presented here,
we also did the same simulation survey
for the cases of equal mass.
We confirm that there is no significant difference on the total mass of escaping stars.
However, the mass distributions would be different for radii larger than $r_{70}$
(70\% Lagrangian radius) in these two models.

Observationally, the initial kinetic energy of the whole star cluster is not easy to estimate.
On the other hand, the cluster-cloud mass ratio is more promising and should warrant
further studies.

\acknowledgments
We thank the referee for valuable suggestions to an earlier version of the paper.
This work was supported in part by the National Science Council, Taiwan
under the grants NSC-96-2112-M-008-014-MY3.


\begin{thebibliography}{}

\bibitem[Aarseth(2001)]{aar01}
  Aarseth S.J., 2001, New Astronomy, 6, 277
\bibitem[Baumgardt \& Kroupa(2007)]{bau07}
  Baumgardt H., Kroupa P., 2007, \mnras, 380, 1589
\bibitem[Bastian et al.(2008)]{bas08}
  Bastian N., Gieles M., Goodwin S. P., Trancho G., Smith L. J.,
  Konstantopoulos I., Efremov Yu., 2008, MNRAS, 389, 223
\bibitem[Bastian \& Goodwin(2006)]{bas06}
  Bastian N., Goodwin S.P., 2006, \mnras, 369, L9
\bibitem[Binney \& Tremaine(1987)]{bin87}
  Binney J., Tremaine S., 1987, Galactic Dynamics (Princeton, NJ : Princeton Univ. Press)
\bibitem[Blitz \& Shu(1980)]{bli80}
  Blitz L., Shu F., 1980, \apj, 238, 148
\bibitem[Boily \& Kroupa(2003a)]{boi03a}
  Boily C.M., Kroupa P., 2003a, \mnras, 338, 665
\bibitem[Boily \& Kroupa(2003b)]{boi03b}
  Boily C.M., Kroupa P., 2003b, \mnras, 338, 673
\bibitem[Bonnell et al.(2006)]{bon06}
  Bonnell I. A., Dobbs C. L., Robitaille T. P., Pringle J.E., 2006, \mnras, 365, 37
\bibitem[Chen \& Ko (2008)] {che08}
  Chen H.-C., Ko C.-M., 2008, Astron. Nachr., 329, 1053
\bibitem[Elmegreen(2000)]{elm00}
  Elmegreen B., 2000, \apj, 530, 277
\bibitem[Gieles et al.(2006)]{gie06}
  Gieles M., Portegies Zwart S.F., Baumgardt H., Athanassoula E., Lamers H.J.G.L.M.,
  Sipior M., Leenaarts J., 2006, \mnras, 371, 793
\bibitem[Geyer \& Burkert(2001)]{gey01}
  Geyer M. P., Burkert A., 2001, \mnras, 323, 988
\bibitem[Goodwin(1997)]{goo97}
  Goodwin S.P., 1997, \mnras, 284, 785
\bibitem[Goodwin(2008)]{goo08}
  Goodwin S.P., 2008, to appear in the proceedings of the meeting, ``Young
massive star clusters - Initial conditions and environment'', ed. E. Perez et al.,
Granada, Spain, September 2007 (Springer: Dordrecht)
\bibitem[Goodwin \& Bastian(2006)]{goo06}
  Goodwin S.P., Bastian N., 2006, \mnras, 373, 752
\bibitem[Hartmann et al.(2001)]{har01}
  Hartmann L., Balesteros-Paredes J., Bergin E. A., 2001, \apj, 562, 852
\bibitem[King(1966)]{kin66}
  King I. R., 1966 AJ, 71,64
\bibitem[Kim et al.(2006)]{kim06}
  Kim S. S., Figer D. F., Kudritzki R. P., Najarro F., 2006, \apj, 653, L113
\bibitem[Lada \& Lada(2003)]{lad03}
  Lada C.J., Lada, E.A. 2003, ARA\&A, 41, 57
\bibitem[Lada et al.(1984)]{lad84}
  Lada C.J., Margulis M., Dearborn D., 1984, \apj, 285, 141
\bibitem[Leisawitz et al.(1989)]{lei89}
  Leisawitz D., Bash F. N., Thaddeus P. 1989, \apjs, 70, 731
\bibitem[Parmentier et al.(2008)]{par08}
  Parmentier G., Goodwin S. P., Kroupa P. and Baumgardt H., 2008, \apj, 678, 347
\bibitem[Plummer(1911)]{plu11}
  Plummer H. C., 1911, MNRAS, 71, 460
\bibitem[Portegies Zwart et al.(2007)]{por07}
  Portegies Zwart S., Gaburov E., Chen H.-C., G\"{u}rkan M. A., 2007, \mnras, 378, 29
\bibitem[Salpeter(1955)]{sal55}
  Salpeter E.E., 1955, \apj, 121, 161
\bibitem[Spitzer(1958)]{spi58a}
  Spitzer L., 1958, ApJ, 127, 544
\bibitem[Spitzer \& H\"{a}rm(1958)]{spi58b}
  Spitzer L., H\"{a}rm R., 1958, 127, 17.
\bibitem[Tutukov(1978)]{tut78}
  Tutukov A. V., 1978, A\&A 70, 57
\bibitem[Williams \& McKee(1997)]{wil97}
  Williams J., McKee C.F., 1997, \apj, 476, 166

\end{thebibliography}
\end{document}